\documentclass[%
 aip,
 amsmath,amssymb,
 reprint,%
]{revtex4-1}

\usepackage{graphicx}
\usepackage{dcolumn}
\usepackage{bm}

\usepackage[utf8]{inputenc}
\usepackage[T1]{fontenc}
\usepackage{mathptmx}
\usepackage{etoolbox}

\usepackage{array}
\usepackage{hyperref}
\usepackage{tikz}
\usepackage{float}
\usepackage{graphicx}
\usepackage[english]{babel}
\usepackage{graphics}
\usepackage{xcolor}
\usepackage{amsmath,amsfonts,amssymb}
\usepackage{soul}

\usepackage{ragged2e}
\usepackage{subcaption}
\newcommand*{\patchcaption}{\justifying\small}
\usepackage{caption}
\newcommand{\comment}[1]{}

\makeatletter
\def\@email#1#2{%
 \endgroup
 \patchcmd{\titleblock@produce}
  {\frontmatter@RRAPformat}
  {\frontmatter@RRAPformat{\produce@RRAP{*#1\href{mailto:#2}{#2}}}\frontmatter@RRAPformat}
  {}{}
}%
\makeatother
\begin{document}

\preprint{AIP/123-QED}

\title[Optimization of electrode morphology and bubble transport for efficient catalysis]{Enhancement of bubble transport by tuning catalyst morphology}
\title[]{Optimized catalyst morphology leads to enhanced bubble transport}
\title{Enhancement of bubble transport in porous electrodes and catalysts}

\author{Thomas Scheel}
\affiliation{Helmholtz Institute Erlangen-N\"urnberg for Renewable Energy, Forschungszentrum J\"ulich, Cauerstr.\,1, D-91058 Erlangen, Germany}
\affiliation{Department of Physics, Friedrich-Alexander-Universit\"at Erlangen-N\"urnberg, Cauerstr.\,1, D-91058 Erlangen, Germany} 

\author{Paolo Malgaretti}
\affiliation{Helmholtz Institute Erlangen-N\"urnberg for Renewable Energy, Forschungszentrum J\"ulich, Cauerstr.\,1, D-91058 Erlangen, Germany}

\author{Jens Harting}
\affiliation{Helmholtz Institute Erlangen-N\"urnberg for Renewable Energy, Forschungszentrum J\"ulich, Cauerstr.\,1, D-91058 Erlangen, Germany}
\affiliation{Department of Chemical and Biological Engineering and Department of Physics, Friedrich-Alexander-Universit\"at Erlangen-N\"urnberg, Cauerstr.\,1, D-91058 Erlangen, Germany} 

\date{\today}

\begin{abstract}
We investigate the formation and transport of gas bubbles across a model porous catalyst/electrode using lattice Boltzmann simulations. This approach enables us to
systematically examine the influence of a wide range of morphologies, flow velocities, and reaction rates on the efficiency of gas production. By exploring these parameters, we identify critical parameter combinations that significantly contribute to an enhanced yield of gas output. Our simulations reveal the existence of an optimal pore geometry for which the product output is maximized. Intriguingly, we also observe that lower flow velocities improve gas production by leveraging on coalescence-induced bubble detachment from the catalyst.
\end{abstract}

\maketitle

\section{Introduction}
The formation and transport of gas bubbles play a pivotal role in a wide range of catalytic reactors and electrolyzers. Many of these applications utilize porous materials, either as catalyst supports or as electrodes themselves~\cite{Haobo2017,Shifa2019,Zhang-Etzold2020,shen_catalytic_2022,McLaughlin-Thiele2023}.
Therefore, it is crucial to optimize the morphology of the porous material to enhance the efficient transport of both reactants and products, aiming at maximizing the overall chemical yield~\cite{Deutschmann_book,detmann_modeling_2021,Bierling-Thiele2023}. 
The case where reactants are in liquid phase and products in gas phase is particularly challenging due to the formation of bubbles, which exhibit dynamics that can be  sensitive to the morphology of the porous materials~\cite{Angulo2020,Schlueter2021,Solymosi2022,Sangtam2023}.

A prominent example for a chemical reaction forming gas bubbles in a liquid phase is the generation of oxygen and hydrogen in an electrolyzer. This process holds significant importance as hydrogen, produced by renewable energy, is a promising alternative to fossil fuels and is expected to play a key role in the transition to a carbon-free economy~\cite{IPMHSMERCBSGSMWVLABNGRSDDHSUWRHFKHG23c}. Nevertheless, the current hydrogen production process is still energy-intensive, and further improvements are necessary to increase its efficiency. Hence, ongoing research focuses on advancements in the development of efficient catalysts to accelerate reaction rates. Building on this, in particular, where the boundaries of material optimization are reached, the search for an optimal electrode morphology takes precedence. However, the design of an efficient electrode microstructure faces a significant challenge: on the one hand maximizing the electrode's surface area is essential to accommodate a large number of gas evolution reactions, while on the other hand, the electrode's morphology must ensure excellent permeability to facilitate the unimpeded transport of gas bubbles away from the electrode's reactive sites to prevent clogging and reduce Ohmic resistances. 

Numerous studies have explored the impact of the porous electrode structure on gas bubble nucleation, growth, detachment and transport aiming at understanding how these factors influence the efficiency of electrolysis~\cite{vanderLinde-Rivas2017, Kou-Li2020, Duhar-Colin2006,Fernandez-Moebius2014,Ito-Yoshida2013,Yuan-Zhan2023}. 
As gas bubble nucleation primarily occurs at the electrode's surface, a common approach to enhance gas output is the maximization of electrode surface area. To achieve this, electrodes are frequently constructed from porous materials, ranging from nanotubes~\cite{Meng-Zhai2017,Assaud-Bachmann2015,Inoue-Kawamura2023}, nanowires~\cite{Xie-Zhang2021,Zhang-Zhao2023}, nanorods~\cite{Huang-Xiong2018,Cen-Shen2021} or nanospheres~\cite{Wang-Huang2022,Zheng-Zhou2022,Bi-Wang2022} to microfibers~\cite{Gu-Nam2017,Yang-Wiley2020} and macroporous foams~\cite{Chan-Li2013,Rocha-Proost2022}. 
These structures cover various length scales, from the nano- to millimeter-scale, combining the advantages of high electrochemical activity and high permeability \cite{Fang-Ho2017,Li-Lu2022}. After nucleation, the bubbles grow by further absorbing dissolved gas via diffusion before they eventually coalesce with neighboring bubbles~\cite{Higuera2022}. This process continues until either the entire reactive site is covered by a gas bubble causing the gas evolution reaction to cease due to the depletion of reactants. Or, alternatively, if the surface tension force attaching the bubble to the electrode is weaker than buoyancy and/or than the viscous stress force exerted on the bubble by a surrounding flow, the gas bubble detaches and is advected away~\cite{Duhar-Colin2006}. In terms of optimizing the electrode efficiency, it is crucial to prevent the formation of bubbles blocking the reactive sites~\cite{Kou-Li2020} and instead to promote their effective removal. Therefore, a frequently utilized approach for improving bubble removal, especially in flow cells, is to apply a flow field in parallel with or even through the porous electrode structure~\cite{Gillespie-Kriek2015,Pletcher-Brown2018,Yang-Wiley2020,Abdelghani-Idrissi-Colin2021}. Another strategy to improve gas bubble detachment encompasses the tuning of electrode wettability, which includes the design of electrode microstructures that promote high contact angles~\cite{Khorasani-Wilkinson2017,Sakuma-Matsushima2014,Meng-Zhai2017,Lu-Jiang2014,Jeon-Ryu2020}. By doing so, the contact area between the gas bubble and the electrode is reduced, which diminishes the surface tension force binding the gas bubble to the electrode and facilitates the detachment of the gas bubble.

Once gas bubbles detach from the surface, the focus shifts to their efficient transport within the porous electrode structure~\cite{Arbabi-Bazylak2016,Lopata-Shimpalee2020,Kim-Jung2020,Li2022,Rocha-Proost2022}. To explore the impact of geometry on transport efficiency, the electrode microstructure is frequently assessed using averaged parameters like porosity, tortuosity, and permeability. These parameters serve as critical factors in the optimization of the gas transport~\cite{Espinoza-Sunden2017}. Nevertheless, comparatively little attention has been devoted to understanding the influence of the morphology of the vacant electrode space, where gas bubble transport occurs, on electrolytic efficiency~\cite{Kim-Jung2020}. This is attributed not least to the high computational demands that have so far prevented a systematic optimization of the 3-dimensional porous electrode structure.

Considering all of these aspects, it becomes evident that an increased surface area, along with a higher reaction rate, enhances gas production only if the generated gas can be transported away from the electrode fast enough to prevent gas bubbles obstructing the reactive surface. An optimized electrode structure has to account for both, a large electrode surface area and at the same time a high permeability, but their intricate interdependency makes simultaneous optimization challenging. This challenge arises from the fact that achieving a larger surface area often results in increased resistance to flux whereas conversely, a planar geometry with high permeability and low tortuosity tends to exhibit a lower surface area. Adding to this complexity is the involvement of length scales spanning several orders of magnitude, making the optimization a very difficult task. Until today, it is not clear how the optimal electrode morphology should be designed. 

\begin{figure}[H]
\begin{center}
    \includegraphics[width=1.0 \columnwidth]{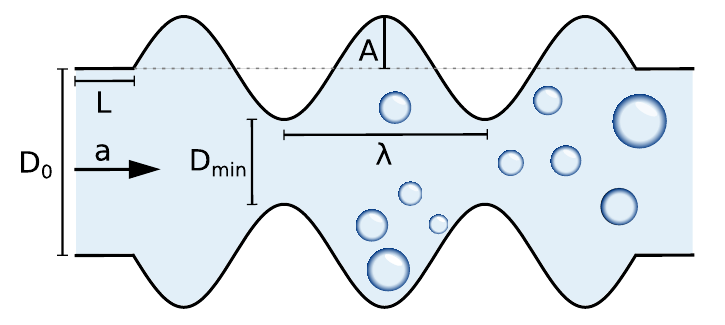} 
    \caption{\patchcaption Schematic representation of the porous model structure used in our simulations.}
    \label{fig:unitcell_sketch}
\end{center}
\end{figure}

To address this challenge we employ the lattice Boltzmann method (LBM) and simulate the electrocatalytic process using a sinusoidal unit cell as a model of the porous electrode or support structure Fig.~\ref{fig:unitcell_sketch}. 
The sinusoidal pore model simplifies the flow geometry to a few essential parameters, enabling a systematic investigation of the impact of morphology, flow velocities, and reaction rates on the catalytic output. This approach allows us to optimize relevant system parameters and identify critical parameter combinations that significantly contribute to an enhanced yield of produced gas.

The subsequent sections of this paper are organized as follows: In Section II we outline our numerical methodology and simulation conditions. Section III comprises a summary of our simulation results, investigating the impact of diverse pore geometries, flow velocities, and reaction rates on catalytic efficiency. The final section provides conclusions and a brief outlook on future research directions.

\section{Numerical method and simulation conditions}

The simulation of the electrolysis process requires an algorithm with the capability to accurately model the intricate interplay of phenomena such as phase separation and chemical reactions with hydrodynamics. We have selected the lattice Boltzmann algorithm as our model of choice, as it excels in meeting these requirements and it has recently been used in similar studies~\cite{falcucci_mapping_2016,Li2022}.

We conduct our lattice Boltzmann simulations on a three-dimensional lattice with 19 discrete velocities (D3Q19)~\cite{Benzi-Vergassola1992}. The evolution of the discrete distribution function $f_i^k(\vec{x},t)$ for each fluid component $k$ is described by the lattice Boltzmann equation
\begin{equation}
f_i^k(\vec{x}+\vec{c}_i \Delta t, t + \Delta t) = f_i^k(\vec{x},t) + \Omega_i^k(\vec{x},t),
\end{equation}
where $\Omega_i^k$ is the collision operator, $i=1,...,19$ specifies the lattice direction and $k \in \{1,2\}$ the fluid component. For clarity without loss of generality, we set the time step $\Delta t=1$, the lattice constant $\Delta x=1$ and the unit mass $m_0=1$ for the subsequent analysis. The fluid density $\rho^k$ is derived from the zeroth moment of the distribution function
\begin{equation}
\rho^k(\vec{x},t) = \sum_i f_i^k(\vec{x},t),
\end{equation}
and  the macroscopic fluid velocity  $\vec{u}^k(\vec{x},t)$ from the first moment of the distribution function
\begin{equation}
\vec{u}^k(\vec{x},t) = \frac{ \sum_i f_i^k(\vec{x},t) \, \vec{c}_i}{\rho^k(\vec{x},t)}.
\end{equation}
Phase separation is implemented using the color gradient method (CG) which introduces an interaction between different fluid components achieving phase separation in three steps \cite{Gunstensen1991,Leclaire-Reggio-Trepanier2013,Montessori-Succi2018,Karun2022}. In the first step, the direction of steepest increase in the density of the respective fluid component (color gradient) is calculated 
\begin{equation}
\vec{F}^k(\vec{x},t) = \nabla \left( \frac{\rho^\zeta(\vec{x},t) - \rho^\xi(\vec{x},t)}{\rho^\zeta(\vec{x},t) + \rho^\xi(\vec{x},t)} \right),
\end{equation}
where $\zeta,\xi \in \{1,2\} \;\; \textrm{for a two component fluid and} \;\; \zeta>\xi$.

In the subsequent perturbation step, populations collinear to the gradient of the density field of the respective fluid component $k$ are increased, whereas populations perpendicular to it are diminished, resulting in the emergence of surface tension:

\begin{equation}
\left( \Omega_i^k \right)^\mathrm{pert} f_i^k (\vec{x},t) = f_i^k(\vec{x},t) + \frac{A_k}{2} |\vec{F}^k(\vec{x},t)| \left( w_i \cos^2(\phi_i^k) - B_i\right),
\label{CG_pert}
\end{equation} 
with $w_i$ being the lattice weights
\begin{equation}
w_i=\left\{%
\begin{array}{ll}
    1/3 & i=1 \\
    1/18 & i=2,...\,,7 \\
    1/36 &i=8,...\,,19 \\
\end{array}%
\right.
\end{equation}
and $\phi_i^k$ the angle between the color gradient $\vec{F}^k$ and the lattice direction $\vec{c}_i$. The strength of the surface tension is controlled using the free parameter $A_k$, while $B_i$ is selected to ensure mass conservation:
\begin{equation}
 B_i=\left\{%
\begin{array}{ll}
    -2/9& i=1\\
    1/54& i=2,...\,,7\\
    1/27& i=8,...\,,19\\
\end{array}%
\right.
\end{equation}
In the final step, known as the recoloring step, the two phases are separated by distributing the two fluid populations in opposite directions.
\begin{align}
\begin{split}
    \left( \Omega_i^\zeta \right)^\mathrm{recol} f_i (\vec{x},t) &= \frac{\rho^\zeta}{\rho} f_i(\vec{x},t)\\  
&+ \beta \frac{\rho^\zeta \rho^\xi}{\rho^2} \cos(\phi_i) \sum_{k=\zeta, \xi} f_i^{k,eq}(\vec{x},t)(\rho^k,0),
\end{split}
\label{recoloring}
\end{align}
where $\beta$ controls the interface thickness ($\beta=0.99$ in all our simulations to obtain sharp interfaces), $f_i=\sum_k f_i^k$. The local equilibrium distribution $f_i^{k,eq}$ is derived from a Taylor expansion of the Maxwell-Boltzmann distribution to the second order
\begin{equation}
f_i^{k,eq}(\vec{x},t) = \rho^k \left[ \phi_i^k + \varphi_i \bar{\alpha} + w_i \left( \frac{\vec{c}_i \cdot \vec{u} }{c_s^2} + \frac{(\vec{c}_i \cdot \vec{u})^2 }{2 c_s^4} - \frac{\vec{u}^2 }{2 c_s^2} \right) \right],
\end{equation}
with $c_s$ being the lattice speed of sound, $\varphi_i$ a lattice dependent weight and $\bar{\alpha}$ the density weighted average of parameter $\alpha_k$, which sets the equilibrium density for each fluid component \cite{Leclaire-Latt2017}.
The total collision operator of the CG method $\Omega_i^k$ extends the standard Bhatnagar-Gross-Krook (BGK) collision operator \cite{Bhatnagar-Krook1954}
\begin{equation}
\left( \Omega_i^k \right)^\mathrm{BGK} f_i^k(\vec{x},t) = f_i^k(\vec{x},t) - \omega_k \left( f_i^k(\vec{x},t) - f_i^{k,eq}(\vec{x},t) \right).
\end{equation}
The BGK collision operator relaxes the population $f_i^k$ to its local equilibrium with a relaxation rate $\omega_k= 1/\tau_k$, with $\tau_k=5.5$ in our simulations. The relationship between the relaxation time $\tau_k$ and the kinematic viscosity $\nu_k$ of fluid $k$ is derived from the Chapman-Enskog expansion to second order. This relationship is expressed as
\begin{equation}
\nu_k = c_s^2 \left( \tau_k- \frac{1}{2}  \right).
\end{equation}
Finally, the BGK collision operator is extended by incorporating the perturbation and recoloring operators to yield the CG collision operator $\Omega_i^k$,
\begin{equation}
\Omega_i^k= \left( \Omega_i^k \right)^\mathrm{recol}\circ\left( \Omega_i^k \right)^\mathrm{pert} \circ \left( \Omega_i^k \right)^\mathrm{BGK}.
\end{equation}
The CG collision operator is constructed by sequentially applying the BGK, perturbation, and recoloring operators, in this order, and conserves all collisional invariants like mass and total momentum for each fluid component.

We carefully validated our simulation framework by successfully reproducing phenomena such as  Neumann angles, the equation of Young-Laplace and the behavior of oscillating droplets, consistent with the data presented in~\cite{Leclaire-Reggio-Trepanier2011}. Furthermore, we achieved outstanding results in simulating the coalescence of liquid lenses across wide ranges of surface tensions and viscosities, as demonstrated in~\cite{Scheel-Harting2023}. 

Chemical reactions are introduced through reactive boundary conditions at lattice nodes adjacent to the reactive boundary, converting all 19 fluid populations of one fluid species into another governed by the reaction rate $\omega$. To minimize the influence of unequilibrated flow, reactions in our simulation occur solely at the boundaries of the middle pore.
We set up our simulations with periodic boundary conditions to prevent the introduction of artificial boundary effects at the channel's inflow and outflow. However, this choice necessitates the removal of gas bubbles at the domain outlet. To achieve this, we skip the recoloring step of the color gradient method in the vicinity of the outlet and instead assign the total population of the perturbation step to the fluid phase, ensuring a bubble-free fluid at the channel outlet. Nonetheless, we allocate one-quarter of the entire domain at both the inlet and the outlet as flow equilibration sections. These domains serve to eliminate boundary effects and prevent interference from the retransformation of gas bubbles into the fluid phase. To ensure the accuracy of our measurements and mitigate biases caused by transient effects, we start our measurements only after a sufficient number of time steps had elapsed so that the simulation could reach either a limit cycle or a steady state (in the case of clogging).

We choose the amplitude $A$ of the sinusoidal channel boundary, the channel diameter $D_0$, the reaction rate $\omega$, and flow velocity induced by an acceleration $a$ on the fluid as key variables to tune the electrolytic efficiency. Further simulation parameters are selected in such a way that they are as similar as possible to the real parameters of water electrolysis, i.e. we aim for low capillary numbers $Ca=\mu U/\sigma <1$ and low Reynolds numbers $Re = \rho U D/\mu \approx 1$. (The low capillary number accounts for the high surface tension of gas in water compared to its low viscosity and a low Reynolds number sets the size of our pores to millimeter scales and below.)

In our set of simulations we varied the amplitudes between $A\in[0;24]$, the channel diameter in the range of $D\in[31;62]$, the applied acceleration by one order of magnitude $a\in[10^{-5};10^{-4}]$ and the reaction rate between $\omega\in[10^{-5};0.003]$ in lattice units. This results in relatively low Reynolds numbers, $Re \in [0.001 ; 2.9]$ and Capillary numbers, $Ca \in [0.001;1.3]$.

We express our parameters and observables in non-dimensional form (denoted by an asterisk (*) hereafter) to eliminate the dependencies on specific units and physical scales. We choose half of the wavelength of the sinusoidal corrugation $\lambda/2$ (see Fig.~\ref{fig:unitcell_sketch}), as our reference length scale. Consequently, the non-dimensionalized amplitude (pore aspect ratio) $A^* = 2A/\lambda$ represents the ratio of pore depth ($2A$) to pore diameter ($\lambda$), controlling the fluid flow through the pore. Similarly, normalizing the channel diameter with the reference length yields $D_0^*=2D_0/\lambda$. Since our simulations take place in the regime of low Reynolds numbers, where viscous forces dominate over inertial forces, we further choose the viscous time $t_c=\lambda^2/(4 \nu)$ (with $\nu$ representing the kinematic viscosity) as our time scale for non-dimensionalization. Consequently, the non-dimensionalized applied acceleration is given by $a^*= a\lambda^3/(8\nu^2)$ and the non-dimensionalized reaction rate by $\omega^*=\omega t_c$. Further, we chose the time-averaged amount of gas leaving the system $\overline{J}_{gas}$ as observable to measure the system efficiency. It is non-dimensionalized by dividing through the maximum outflow observed in all our simulations $\overline{J}_{gas}^*=\overline{J}_{gas}/\overline{J}^{\,max}_{gas}$.

Finally, it should be noted that our freedom of parameter choice is limited by the simulation method. Spurious currents~\cite{Connington-Lee2012,Leclaire-Reggio-Trepanier2011,Ginzbourg-Adler1995,LKLSNJWVH16} are inherent to most lattice Boltzmann multiphase methods and are generally proportional to the surface tension $\sigma$. To ensure that they do not distort the physical flow field, surface tension forces must be small compared to inertial forces. Conversely, the upper limit of fluid velocities $u$ is constrained by the requirement of our LBM implementation to stay within the limit of low Mach numbers.

\section{Simulation results}
Our investigation focuses on the optimization of relevant system parameters and the identification of critical parameter combinations to enhance the electrocatalytic efficiency of electrolyzers. Specifically, we focus on exploring the effects of pore morphology described by corrugation amplitude $A$ and channel diameter $D_0$, flow velocity within the pore, and reaction rate $\omega$. 

These four parameters play a crucial role in optimizing the efficiency of the catalytic process, as underscored by their key contribution to determining the dimensionless Damköhler number $Da=\omega l/u$, where $\omega$ represents the reaction rate, $u$ is the characteristic velocity in the channel, and $l$ is a characteristic length of the system. The Damköhler number serves as an important descriptor of the process efficiency, which signifies the relative importance of the production time scale compared to the advection time scale. The reaction rate, a key parameter in our study, directly influences the $Da$ number. Similarly, when associating the critical length with the length of the reacting boundary, the corrugation amplitude $A$ also influences $Da$ directly. The flow velocity, as the third factor in the $Da$ number, undergoes an intricate interplay between the applied acceleration and pore morphology, as elucidated by our simulations. 

In the following, we examine the influence of each parameter in detail, also considering how variations in other parameters can alter the observed effects through cross-dependencies.

\subsection{Variation of geometry}
\begin{figure}[h]
\centering
\begin{subfigure}[b]{0.95\linewidth}
  \centering
 \includegraphics[width=\textwidth]{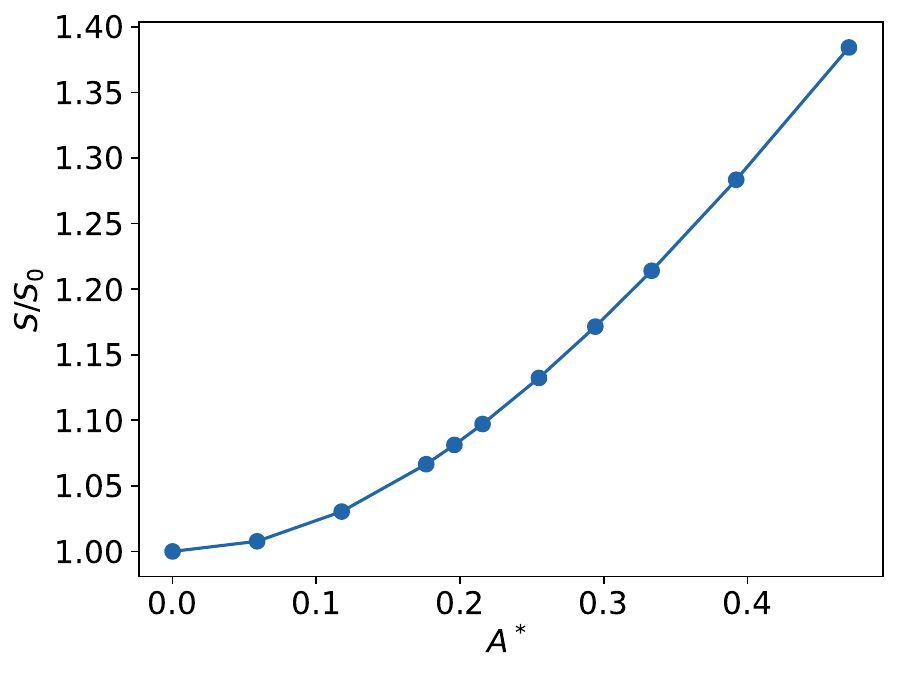}
  \caption{surface area}
\end{subfigure}

\begin{subfigure}[b]{1.0\linewidth}
  \centering
  \includegraphics[width=\textwidth]{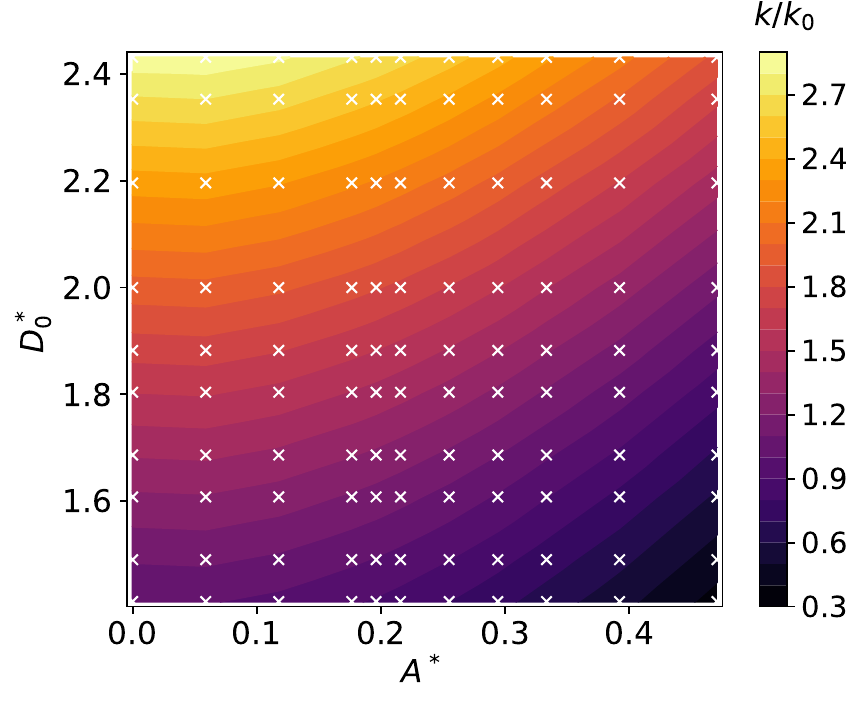}
  \caption{permeability}
\end{subfigure}

\caption{\patchcaption Influence of geometric properties on reacting surface area $S$ normalized by the surface area of the flat channel $S_0$ (a) and permeability $k$ normalized by the permeability of the flat channel (with lowest diameter) $k_0$ (b).}
\label{fig:A-D_SurfaceAndDarcy}
\end{figure}

We commence our investigation by examining how geometry influences gas output. To this end we systematically modify the geometry of the porous structure by varying the pore depth controlled by the amplitude $A$ of the sinusoidal perturbation and by changing the channel diameter $D_0$. An increase in $A$ results in a simultaneous increase of the reactive boundary where gas evolution reactions occur (see Fig.~\ref{fig:A-D_SurfaceAndDarcy}). In our simulation, the reactive surface area $S$ increases by approximately $38\%$ for the highest corrugation amplitude compared to a flat channel. Thus, the geometry with the largest amplitude has the potential to produce the greatest amount of hydrogen.  

However, the full realization of this potential depends on the efficiency of the gas removal process. It is crucial to keep the reactive sites free of reaction products, requiring sufficient permeability of the geometry and, even more critically, adequate advection in proximity to the reactive surface $S$. In a first step, we simulated the flow of a single-phase fluid through the pore geometries and determined their permeabilities using Darcy's law~\cite{NZRHH10,NYLH13}. Taking the permeability of the flat channel with minimal channel diameter $k_0$ as reference, we observe a decrease in permeability by approximately $64\%$ for the maximal amplitude. Furthermore, enlarging the channel diameter resulted in a nearly 3-fold enhancement of permeability.
Combining the effects of surface area and permeability suggests an increase in gas production with increasing amplitude until reaching a maximum. Beyond this point the diminishing permeability becomes predominant, resulting in a decline in gas output. 

\begin{figure}[h]
\begin{center}
    \includegraphics[width=1 \linewidth]{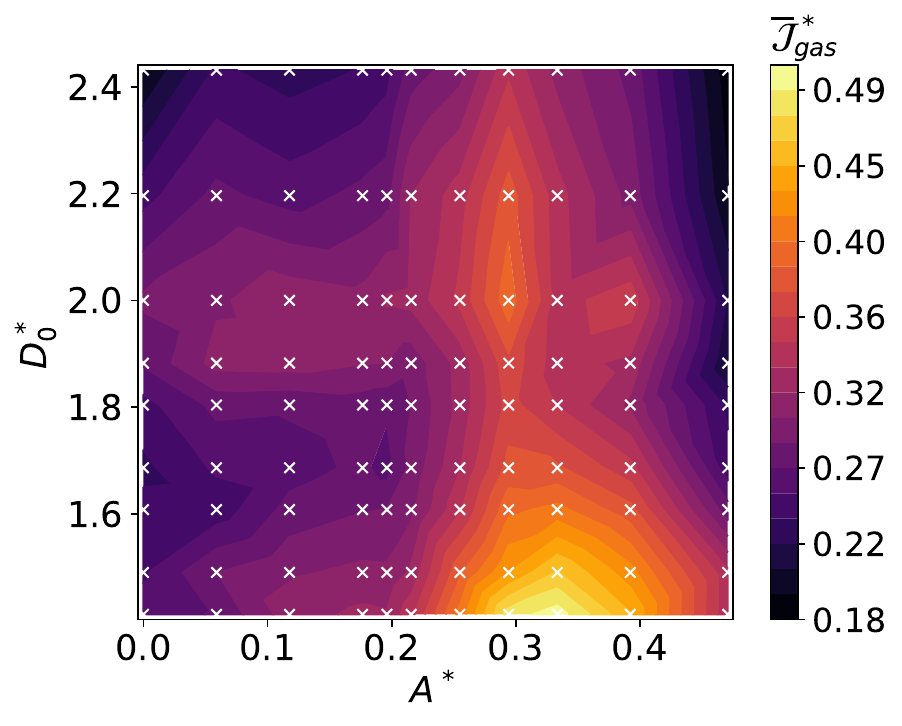} 
    \includegraphics[width=1 \linewidth]{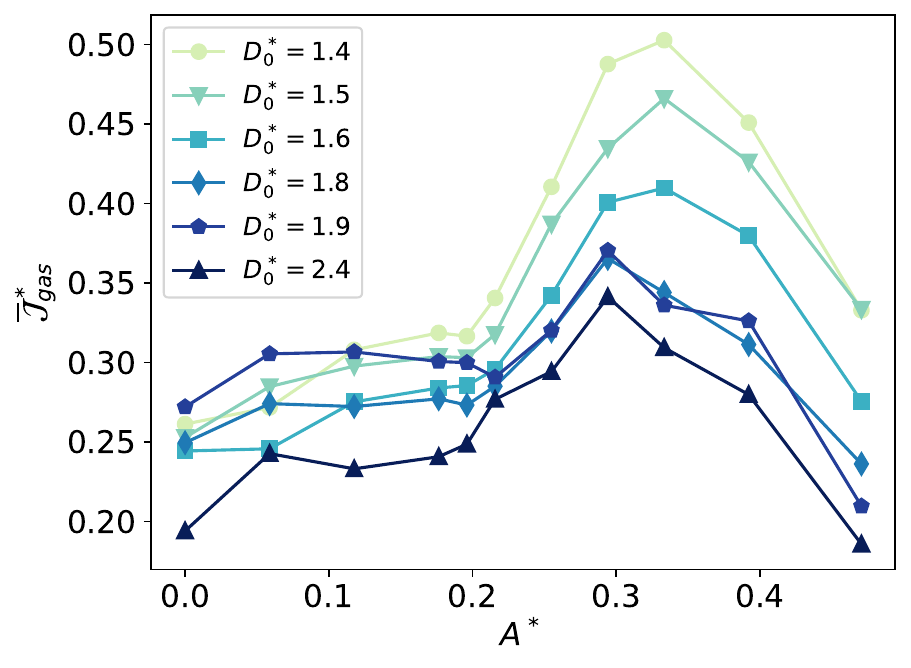} 
    \caption{\patchcaption Mean gas output dependent on channel amplitude $A^*$ and channel diameter $D_0^*$ ($a^*=2.4$, $\omega^*=2.0 \times 10^3$).}
    \label{fig:A-h0_outflow_Gas_mean_2D}
\end{center}
\end{figure}

\begin{figure*}[ht]
\centering
\begin{subfigure}[b]{0.325\linewidth}
  \centering
  \includegraphics[width=\linewidth]{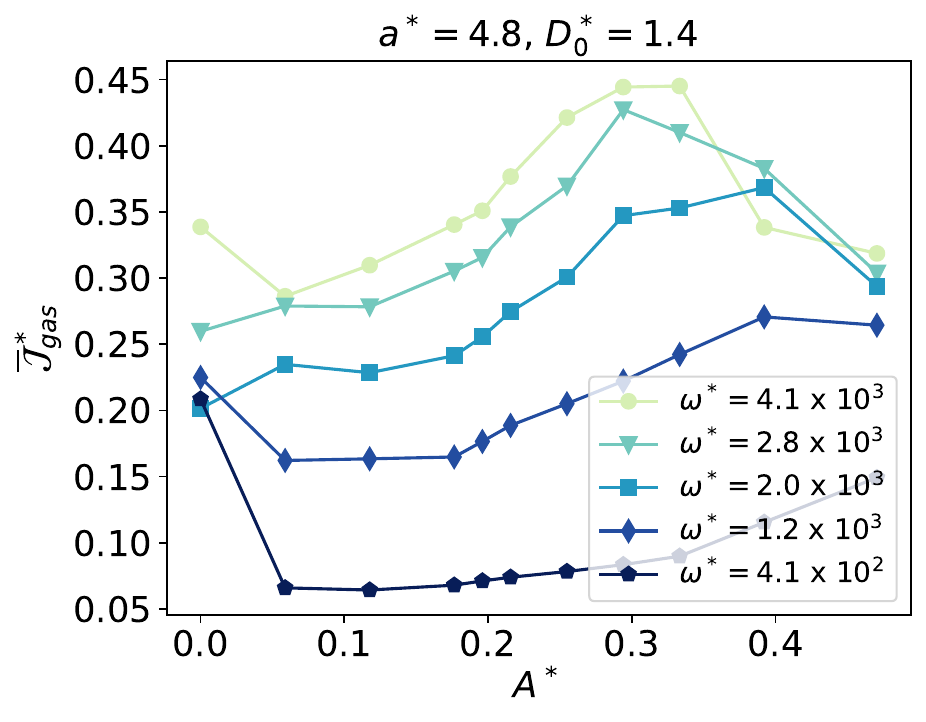}
  \caption{}
  \label{fig:J_A_h0__a}
\end{subfigure}
\begin{subfigure}[b]{0.325\linewidth}
  \centering
  \includegraphics[width=\linewidth]{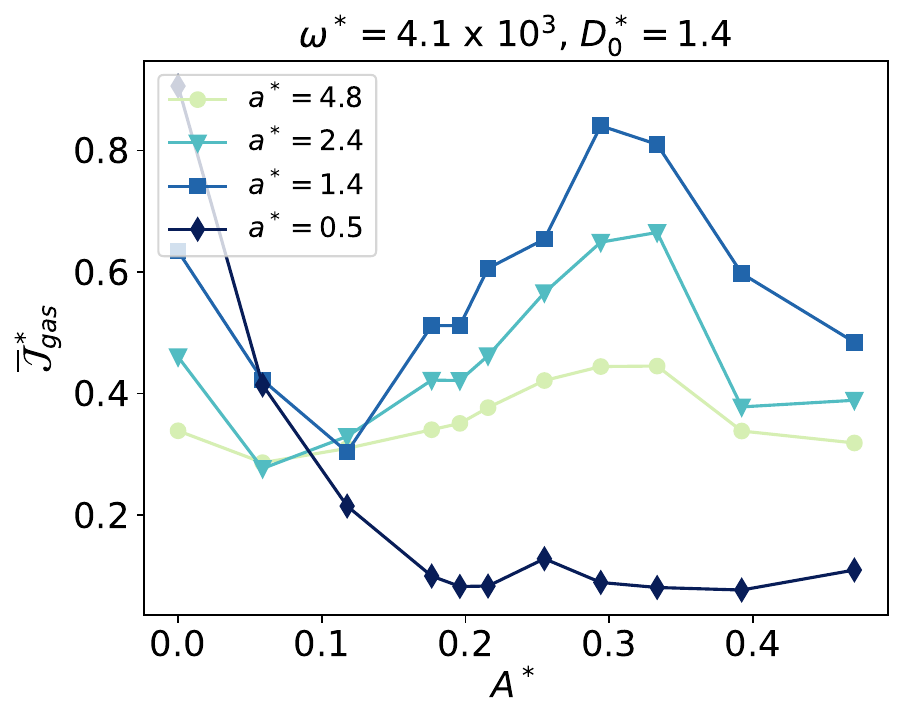}
  \caption{}
  \label{fig:J_A_h0__b}
\end{subfigure}
\begin{subfigure}[b]{0.325\linewidth}
  \centering
  \includegraphics[width=\linewidth]{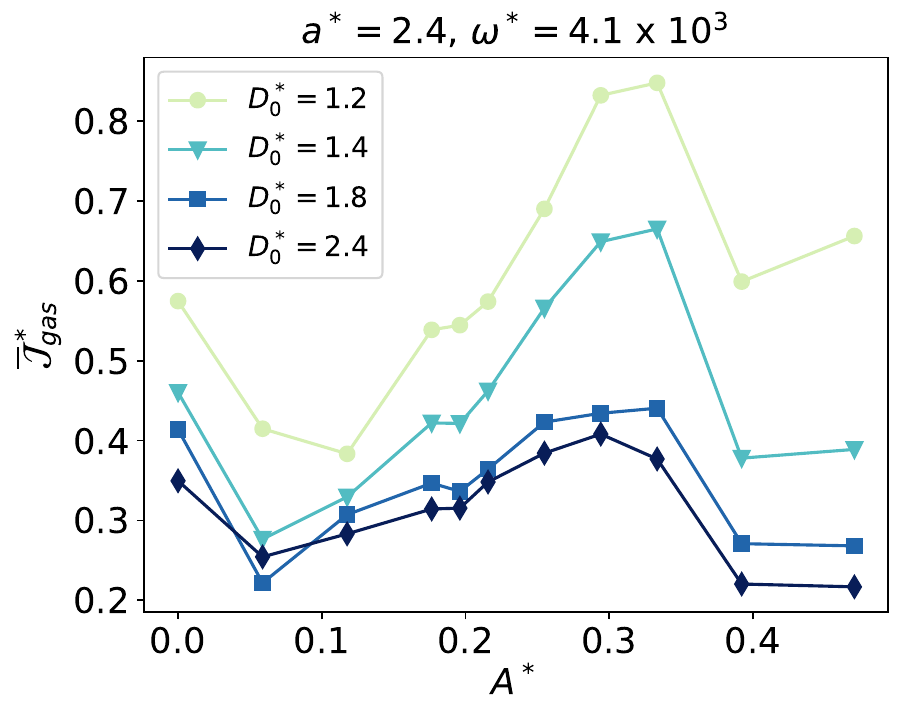}
  \caption{}
  \label{fig:J_A_h0__c}
\end{subfigure}
\begin{subfigure}[b]{0.325\linewidth}
  \centering
  \includegraphics[width=\linewidth]{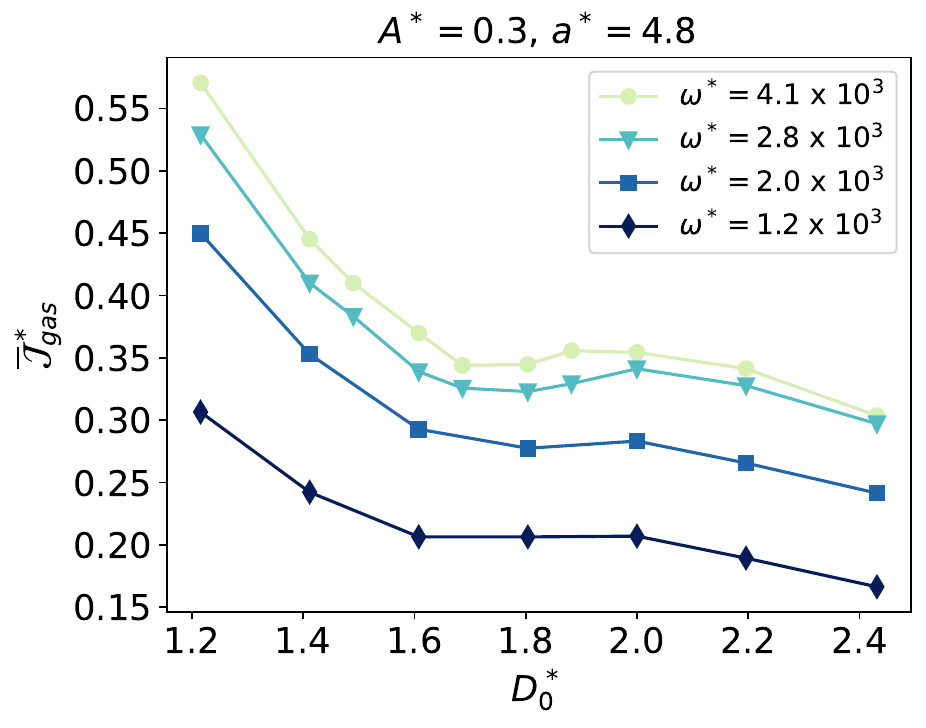}
  \caption{}
  \label{fig:J_A_h0__d}
\end{subfigure}
\begin{subfigure}[b]{0.325\linewidth}
  \centering
  \includegraphics[width=\linewidth]{J_h0_a_nd.pdf}
  \caption{}
  \label{fig:J_A_h0__e}
\end{subfigure}
\begin{subfigure}[b]{0.325\linewidth}
  \centering
  \includegraphics[width=\linewidth]{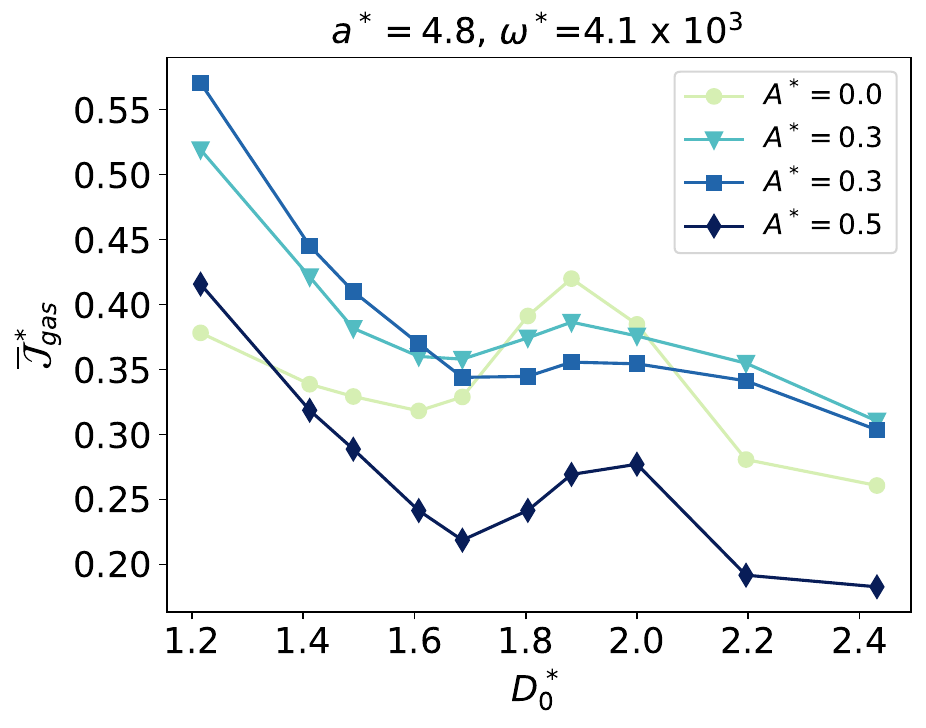}
  \caption{}
  \label{fig:J_A_h0__f}
\end{subfigure}
\caption{\patchcaption Representative one-dimensional slices of the phase space illustrating the dependency of gas output on the corrugation amplitude $A^*$ (top panel) and the channel diameter $D_0^*$ (bottom panel). The left panel varies acceleration, the center panel varies the reaction rate $\omega^*$, and the right panel varies the channel diameter $D_0^*$ or amplitude $A^*$, respectively.}
\label{fig:J_A_h0}
\end{figure*}

\begin{figure}[h]
\centering
\begin{subfigure}[b]{\linewidth}
  \centering
  \includegraphics[width=\linewidth]{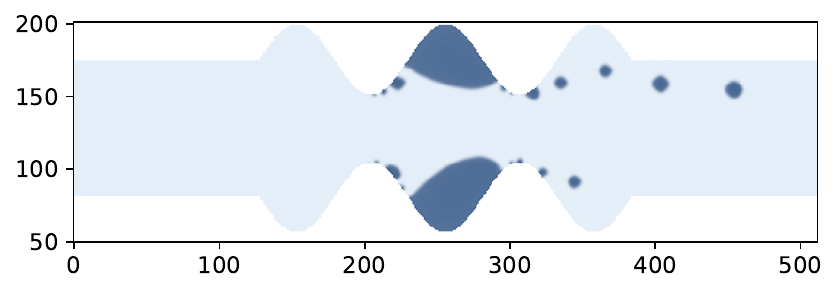}
  \caption{blocked pore}
  \label{fig:blocked_and_clocked__a}
\end{subfigure}

\begin{subfigure}[b]{\linewidth}
  \centering
  \includegraphics[width=\linewidth]{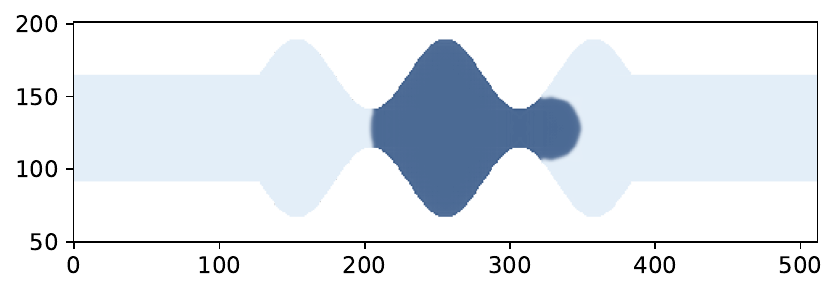}
  \caption{clogged channel}
  \label{fig:blocked_and_clocked__b}
\end{subfigure}

\caption{Representative depictions of a blocked pore (a) and a clogged channel (b).}
\label{fig:blocked_and_clocked}
\end{figure}

We can confirm this expectation through our simulations. As illustrated in Fig.~\ref{fig:J_A_h0} (top panel), a flat channel geometry demonstrates already a good level of gas output for a wide range of accelerations, reaction rates, and channel diameters. This is attributed to the inherent efficiency of a flat geometry, facilitating effective gas transport and minimizing the likelihood of clogging. However, we observe that especially under elevated reaction rates (Fig.~\ref{fig:J_A_h0__a}), accelerations that avoid clogging (Fig.~\ref{fig:J_A_h0__b}) or small channel diameters (Fig.~\ref{fig:J_A_h0__c}) the gas output can be increased by increasing the corrugation amplitude. Under these conditions, the gas output reaches an optimum for a pore aspect ratio $A^*\approx 1/3$. Interestingly, in many cases, the increase in production exceeds the increase in surface area (Fig.~\ref{fig:A-D_SurfaceAndDarcy} and Fig.~\ref{fig:A-h0_outflow_Gas_mean_2D}), indicating that the catalytic area alone is insufficient for predicting the overall production.
However, by increasing amplitudes beyond the optimal threshold, the higher amplitudes extend so deeply into the channel that they impede the efficient transport of gas bubbles, leading to a decrease in gas output. Simultaneously, the increased depth of pores leads to the entrapment and accumulation of gas bubbles, consequently diminishing the number of active reactive sites (Fig.~\ref{fig:blocked_and_clocked__a}). With further increasing amplitudes, the decline in gas output continues until the bubble sizes become comparable to the channel diameter, leading to clogging of the channel with gas bubbles (Fig.~\ref{fig:J_A_h0__b}, diamond symbol and Fig.~\ref{fig:blocked_and_clocked__b}).

Concerning the relationship between gas output and channel diameter, we observe that small channel diameters are generally favorable for most of the parameter space (Fig.~\ref{fig:J_A_h0}, bottom panel). Nevertheless, it is crucial to ensure that the channel diameter is sufficiently large compared to the bubble diameter to prevent channel clogging (Fig.~\ref{fig:J_A_h0__e}, diamond symbol). Depending on the flow velocity in the channel, we observe that channel clogging starts when bubble radii exceed approximately 0.1 to 0.5 times the minimal diameter of the channel.  

Furthermore, we frequently observe a second local optimum at medium channel diameters, which, in special cases like in a flat channel with high reaction rates and high acceleration (Fig.~\ref{fig:J_A_h0__f}, sphere symbol), becomes the global optimum for this parameter combination.

Experiments also indicate that enhancing the electrode surface through corrugation, such as using a 3D-printed NiFe-LDH pyramid structure, results in higher OER performance compared to a flat electrode~\cite{Ahn-Seol2022}. Additionally, experiments confirm that the electrode with the best balance between surface area and bubble permeability exhibits the best performance~\cite{Yang-Wiley2020}.

Taking all factors into consideration, we can derive generic design principles for the geometric configuration of the channel from our simulation results. Firstly, small yet non-clogging channel diameters prove to be favorable. As for the amplitude, our second geometric parameter, either a flat channel or moderate corrugation amplitudes are advantageous, depending on the other non-geometric parameters, i.e., the acceleration and reaction rate. We will discuss the influence of these parameters in the following.

\subsection{Variation of applied pressure drop}

We now focus on elucidating the relationship between the applied pressure drop, modeled via a constant applied acceleration on the fluids, and output of reaction products $\overline{J}_{gas}$. For a given channel height, we can vary the acceleration by approximately one order of magnitude. At lower accelerations, we encounter limitations due to clogging, while at higher accelerations that result in flow velocities approaching $1/3$ of the lattice speed, stability issues with the lattice Boltzmann algorithm become the primary constraint. 

\begin{figure}[ht]
\centering
\begin{subfigure}[b]{\linewidth}
  \centering
  \includegraphics[width=\linewidth]{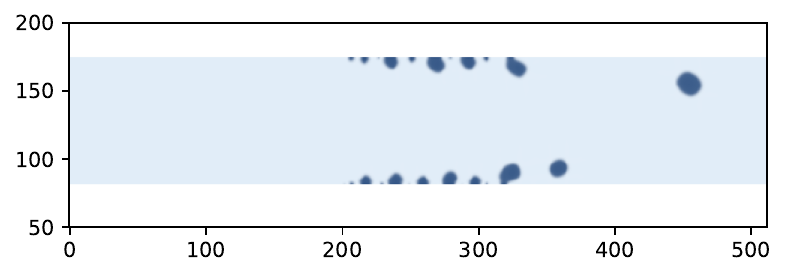}
  \caption{$a^*=3.82$}
\end{subfigure}

\begin{subfigure}[b]{\linewidth}
  \centering
  \includegraphics[width=\linewidth]{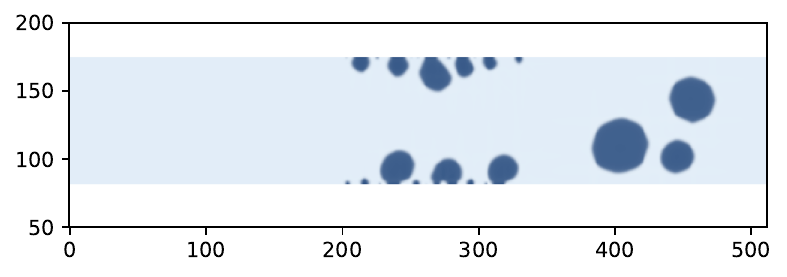}
  \caption{$a^*=0.96$}
\end{subfigure}

\caption{Representative depiction of the impact of accelerations on bubble sizes within a flat channel.}
\label{fig:GasSizesAcc}
\end{figure}

In a simple system with no phase separation and flat geometry, the catalytic output can be enhanced by increasing the velocity within the channel. This improvement continues until a point is reached where all reaction products are promptly transported away, leading to a plateau in the output. However, this line of reasoning is not applicable in situations involving phase separation. In such cases, an intriguing phenomenon arises: reducing the flow velocity within the channel can actually lead to an increase in gas output. While this might initially appear counterintuitive, a closer examination of the simulation data unveils the underlying rationale for this behavior. When flow velocities are decreased, larger bubble radii are observed due to a reduction in drag forces (Fig.~\ref{fig:GasSizesAcc}). 

\begin{figure}[ht]
\centering
\begin{subfigure}[b]{\linewidth}
  \centering
  \includegraphics[width=\linewidth]{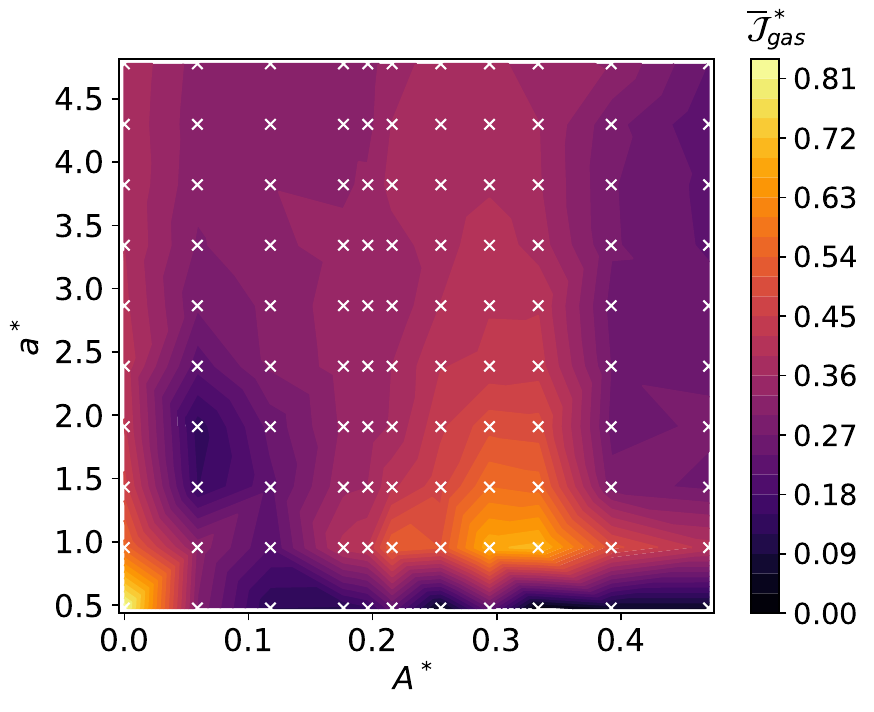}
  \caption{Gas output versus acceleration $a^*$ and amplitudes $A^*$}
  \label{fig:A-acc_outflow_Gas_mean_2D__a}
\end{subfigure}

\begin{subfigure}[b]{\linewidth}
  \centering
  \includegraphics[width=\linewidth]{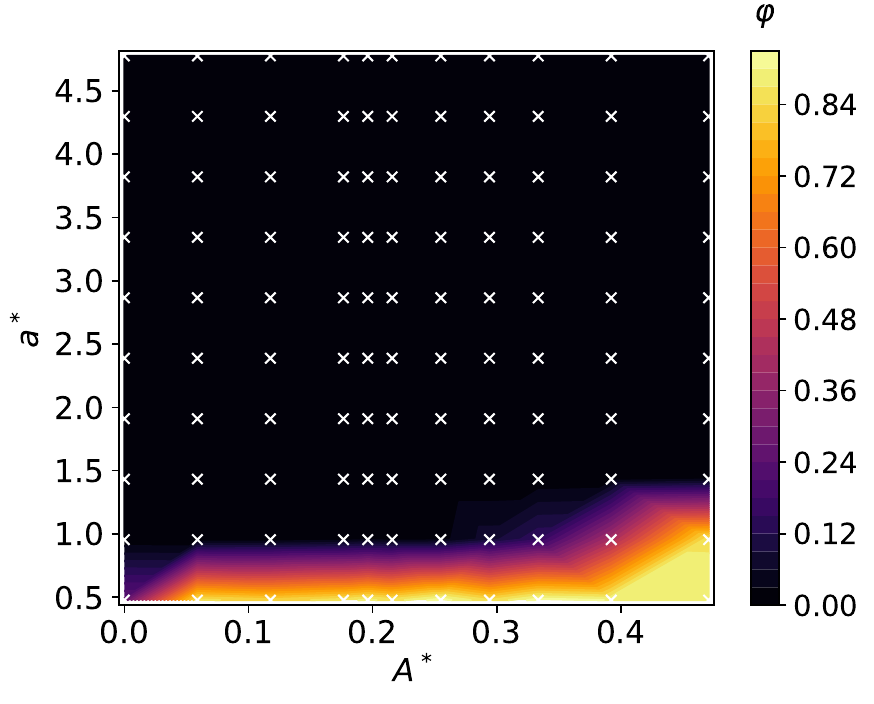}
  \caption{Clogging fraction $\varphi$ versus acceleration $a^*$ and amplitudes $A^*$}
  \label{fig:A-acc_outflow_Gas_mean_2D__b}
\end{subfigure}

\caption{\patchcaption Average gas output (a) and fraction of time that the channel is clogged denoted as $\varphi$ (b) dependent on channel amplitude $A$ and applied pressure gradient (acceleration). All data was simulated with a channel height of $D_0^*=1.8$ and a reaction rate of $\omega^*=4.1 \times 10^3$.}
\label{fig:A-acc_outflow_Gas_mean_2D}
\end{figure}

By further reducing the flow velocity, these bubbles eventually grow to a size where they come into contact with neighboring bubbles. Consequently, these larger bubbles exhibit a higher frequency of coalescence. The process of coalescence results in the detachment of bubbles from the reactive boundaries, thus freeing up the reactive sites for subsequent reactions. Experimental evidence also supports the detachment of bubbles through coalescence, as observed in~\cite{Lv-Lohse2021,Iwata-Wang2022,Raza-Golobic2023}. The efficiency is not solely enhanced by freeing reactive sites through coalescence-induced bubble detachment, but is also complemented by the observation that the detached bubbles tend to move toward the center of the flow channel. In this region, flow velocities are higher compared to those near the surface, resulting in enhanced transport efficiency. This also manifests as a consistent trend in our simulations: reducing acceleration is associated with an increase in gas output across all parameter combinations (Fig.~\ref{fig:A-acc_outflow_Gas_mean_2D} and Fig~\ref{fig:phase-space-gasout}). 

However, this trend only persists until a critical threshold is reached, below which the small acceleration leads to channel clogging. Two factors contribute to this phenomenon: firstly, the low drag forces promote the formation of large gas bubbles with diameters comparable to the channel diameter, eventually leading to clogging of the channel and secondly, the low flow velocities are then insufficient to clear blockages. As illustrated in Fig.~\ref{fig:A-acc_outflow_Gas_mean_2D}, clogging can be observed for small accelerations $a^*<0.96$ and corrugated geometries with amplitudes $A^*\geq 0.06$. Hence, the optimal point for maximizing gas production is generally found at accelerations just before clogging occurs. 

\begin{figure*}[ht]
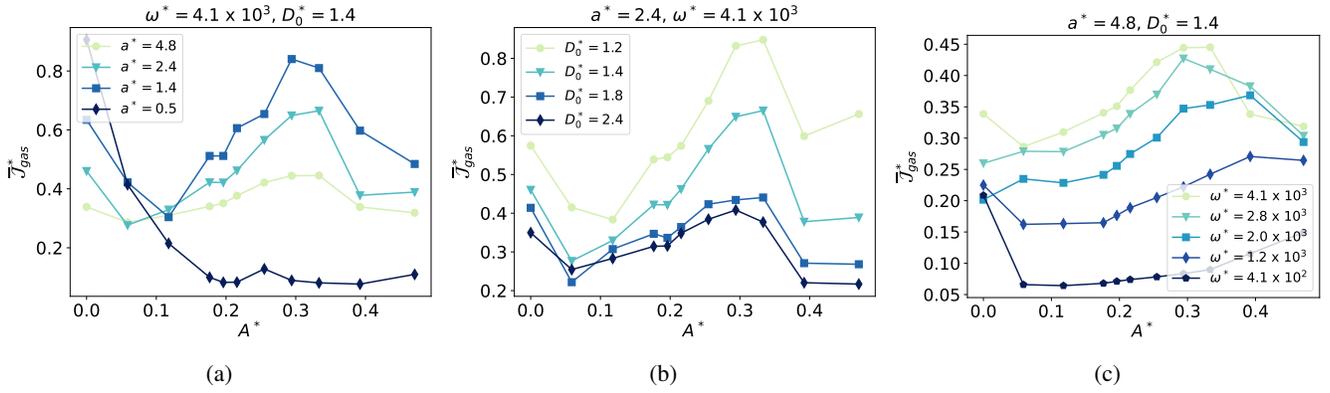

\centering
\begin{subfigure}[b]{0.325\linewidth}
  \centering
  \includegraphics[width=\linewidth]{J_a_A_nd.pdf}
  \caption{}
  \label{fig:J_a__a}
\end{subfigure}
\begin{subfigure}[b]{0.325\linewidth}
  \centering
  \includegraphics[width=\linewidth]{J_a_h0_nd.pdf}
  \caption{}
  \label{fig:J_a__b}
\end{subfigure}
\begin{subfigure}[b]{0.325\linewidth}
  \centering
  \includegraphics[width=\linewidth]{J_a_omega_nd.pdf}
  \caption{}
  \label{fig:J_a__c}
\end{subfigure}
\caption{\patchcaption Representative one-dimensional slices of the phase space, illustrating the variation of gas output in dependence on the acceleration $a^*$ for different values of the amplitude (a), the channel diameter $D_0^*$ (b), and the reaction rate $\omega^*$(c).}
\label{fig:J_a}
\end{figure*}

Referring to Fig.~\ref{fig:J_a}, three general strategies exist to prevent channel clogging under low accelerations. Firstly, the channel diameter can be chosen sufficiently large to avoid clogging at a given reaction rate. Conversely, the reaction rate can be reduced to a level where bubbles are small enough to prevent channel blocking. Both strategies, however, result in a diminished gas output. Notably, the choice of a flat channel geometry stands out as the most favorable configuration for accommodating low accelerations without compromising gas production (Fig.~\ref{fig:J_a__a}, sphere symbol). Additionally, analogous to the simulations with varying channel geometry, we also observe an increase in the gas output for moderate amplitudes in all non-clogging cases when varying the acceleration (Fig.~\ref{fig:A-acc_outflow_Gas_mean_2D__a} and Fig.~\ref{fig:J_a__a}).

\subsection{Variation of reaction rate}

\begin{figure}[ht]
\begin{center}
    \includegraphics[width=1 \linewidth]{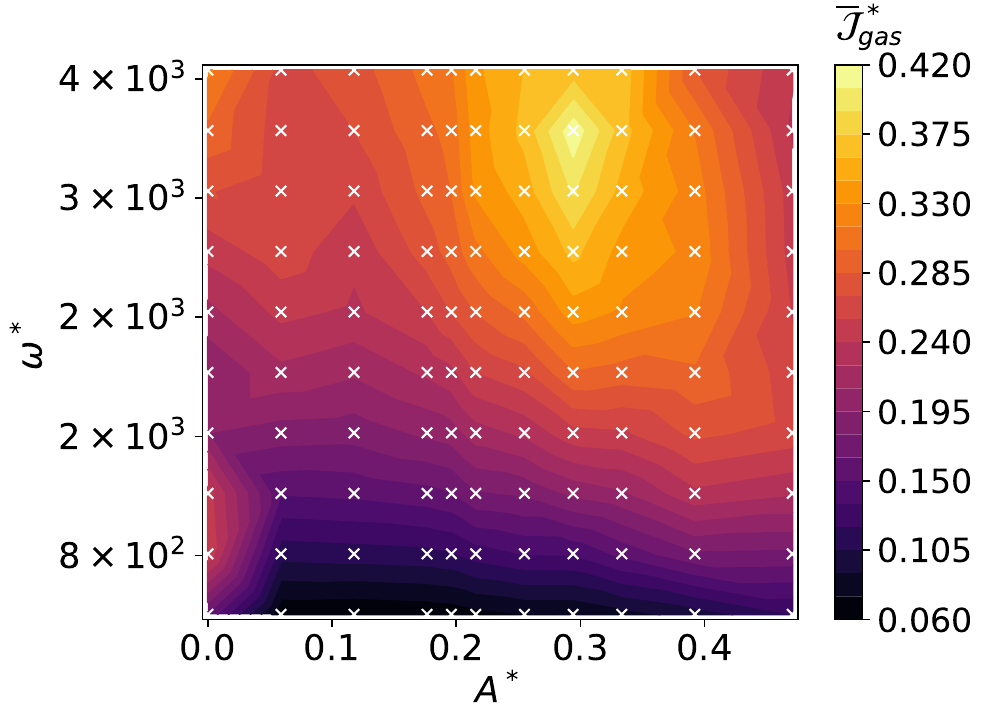} 
    \caption{\patchcaption Mean gas output dependent on channel amplitude $A$ and reaction rate $\omega^*$ ($D_0^*=1.6$, $a^*=4.8$).} 
    \label{fig:A-rr_outflow_Gas_mean_2D}
\end{center}
\end{figure}

\begin{figure*}[ht]
\centering
\begin{subfigure}[b]{0.325\linewidth}
  \centering
  \includegraphics[width=\linewidth]{J_omega_a_nd.pdf}
  \caption{}
  \label{fig:J_omega__a}
\end{subfigure}
\begin{subfigure}[b]{0.325\linewidth}
  \centering
  \includegraphics[width=\linewidth]{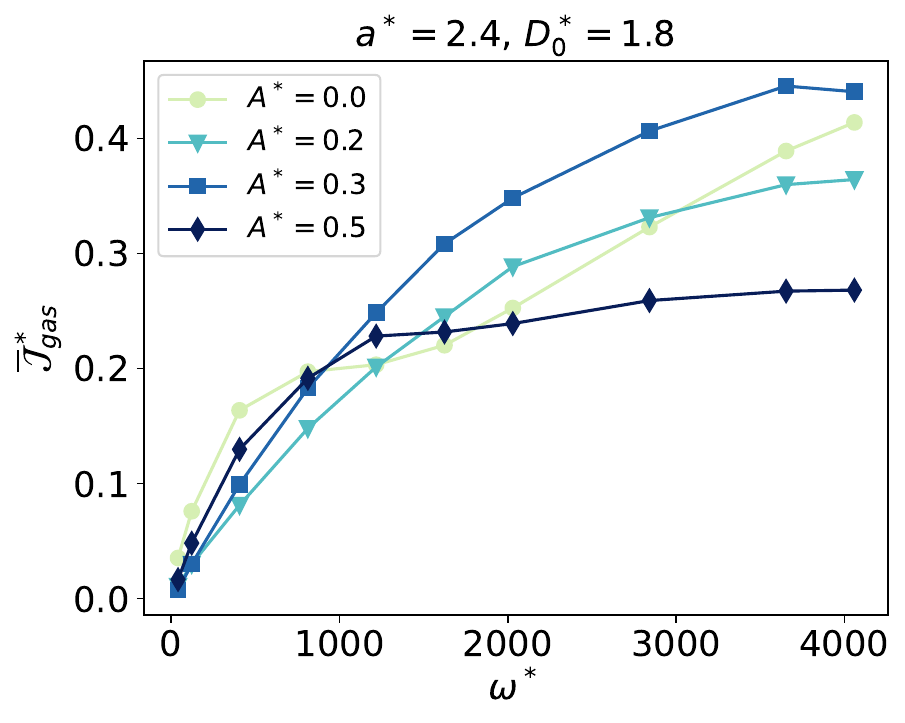}
  \caption{}
  \label{fig:J_omega__b}
\end{subfigure}
\begin{subfigure}[b]{0.325\linewidth}
  \centering
  \includegraphics[width=\linewidth]{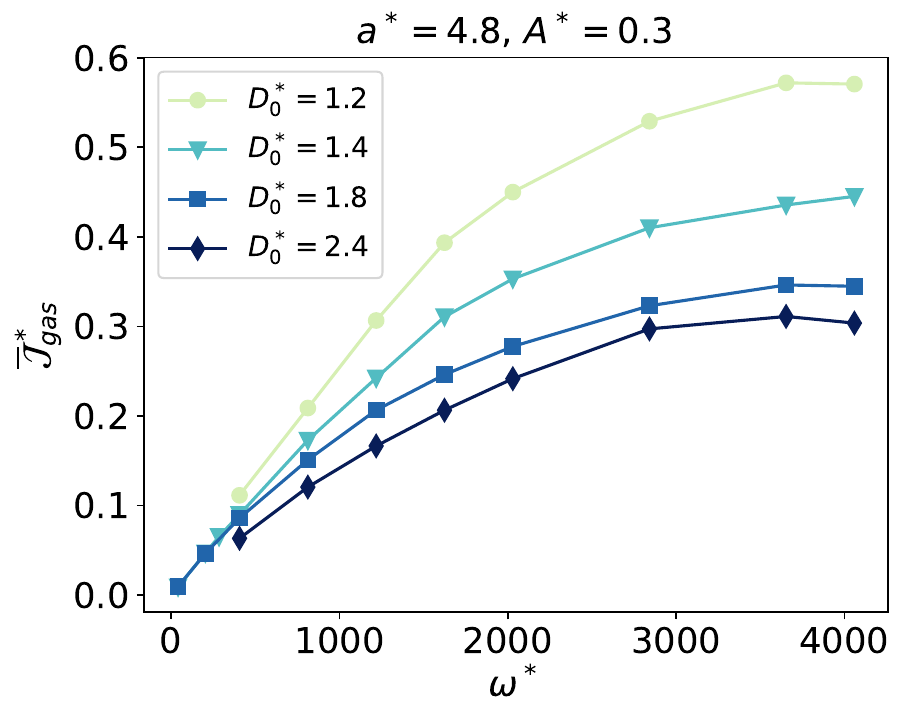}
  \caption{}
  \label{fig:J_omega__c}
\end{subfigure}
\caption{\patchcaption Representative one-dimensional slices of the phase space, illustrating the variation of gas output dependent on the reaction rate $\omega^*$ for different values of acceleration $a^*$ (a), amplitude $A^*$ (b), and the channel diameter $D_0^*$ (c). Channel clogging (a)-diamond symbol and pore blockage (b)-diamond symbal can be observed.}
\label{fig:J_omega}
\end{figure*}

Finally, we explore the impact of the reaction rate $\omega$ on gas output (see Fig.~\ref{fig:A-rr_outflow_Gas_mean_2D}). As expected, enhancing the reaction rate results in a corresponding increase in the gas output. For very high reaction rates the gas output smoothly reaches a plateau (Fig.~\ref{fig:J_omega}), as the reaction rate  becomes so substantial that more gas is produced than can be effectively advected in the channel. Two notable cases arise: firstly, when reaction rate compared to advection becomes too large, resulting in pore blockage (Fig.~\ref{fig:J_omega__b}, diamond symbol), and secondly, when the reaction rate increases to the extent that bubbles with sizes are produced that are comparable to the channel diameter, leading to channel clogging (Fig.~\ref{fig:J_omega__a}, diamond symbol).

In accordance with previous observations, the output increases as the channel amplitude rises with a maximum output at a pore aspect ratio of $A^*\approx 1/3$. Similarly, consistent with previous observations, gas output increases for lower accelerations (Fig.~\ref{fig:J_omega__a}) and smaller channel diameters (Fig.~\ref{fig:J_omega__c}) provided no channel clogging occurs.

\begin{figure*}[ht]
  \centering
  \begin{subfigure}{.3\linewidth}
    \centering
    \includegraphics[width=\linewidth]{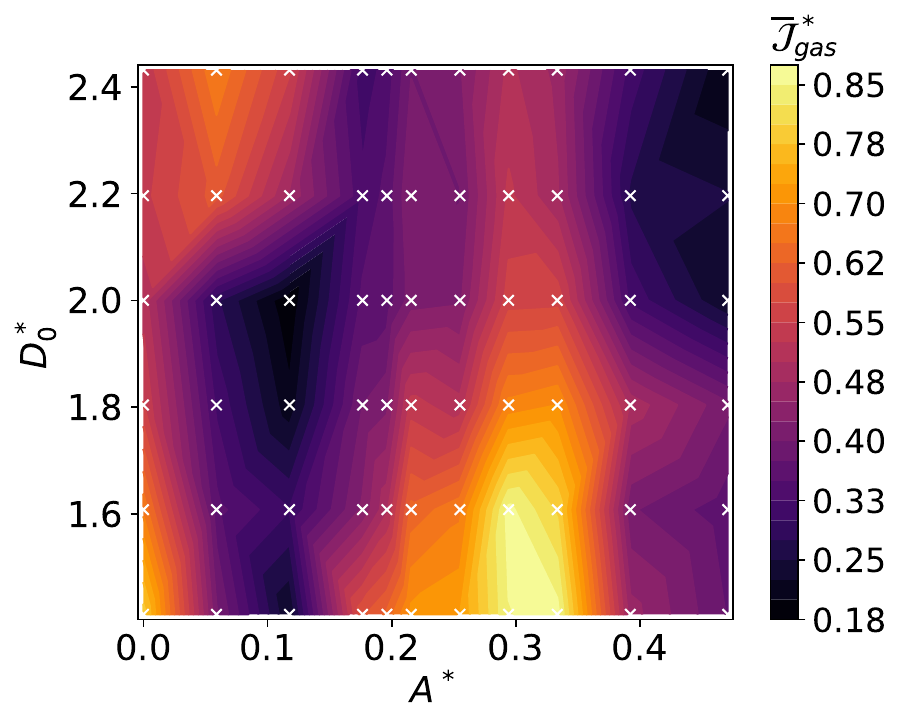}
    \caption{$a^*=1.0$, $\omega^*=4.1 \times 10^3$}
  \end{subfigure}%
  \begin{subfigure}{.3\linewidth}
    \centering
    \includegraphics[width=\linewidth]{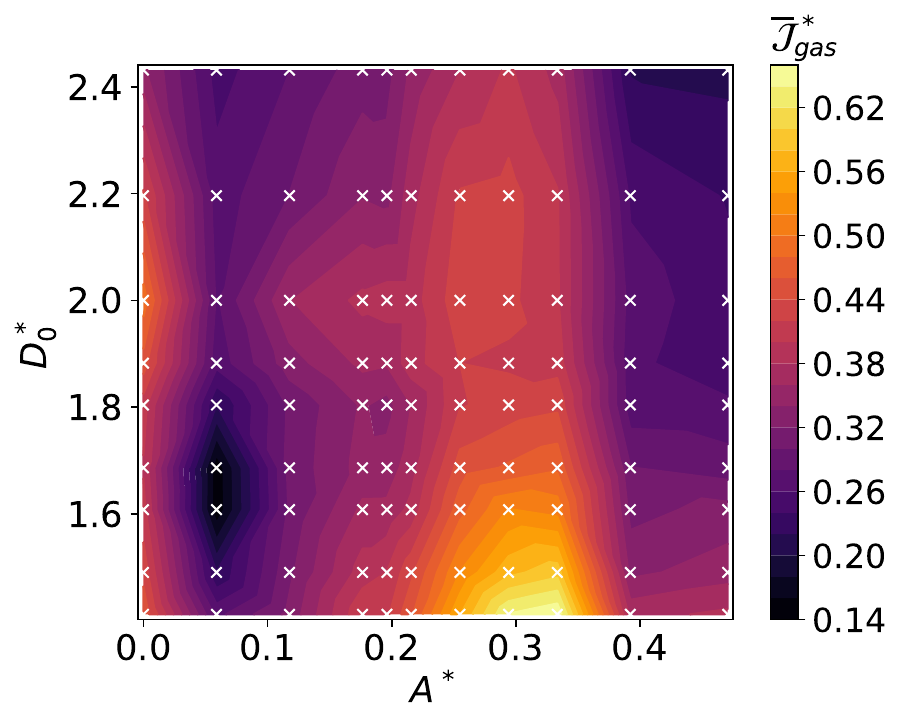}
    \caption{$a^*=2.4$, $\omega^*=4.1 \times 10^3$}
  \end{subfigure}%
  \begin{subfigure}{.3\linewidth}
    \centering
    \includegraphics[width=\linewidth]{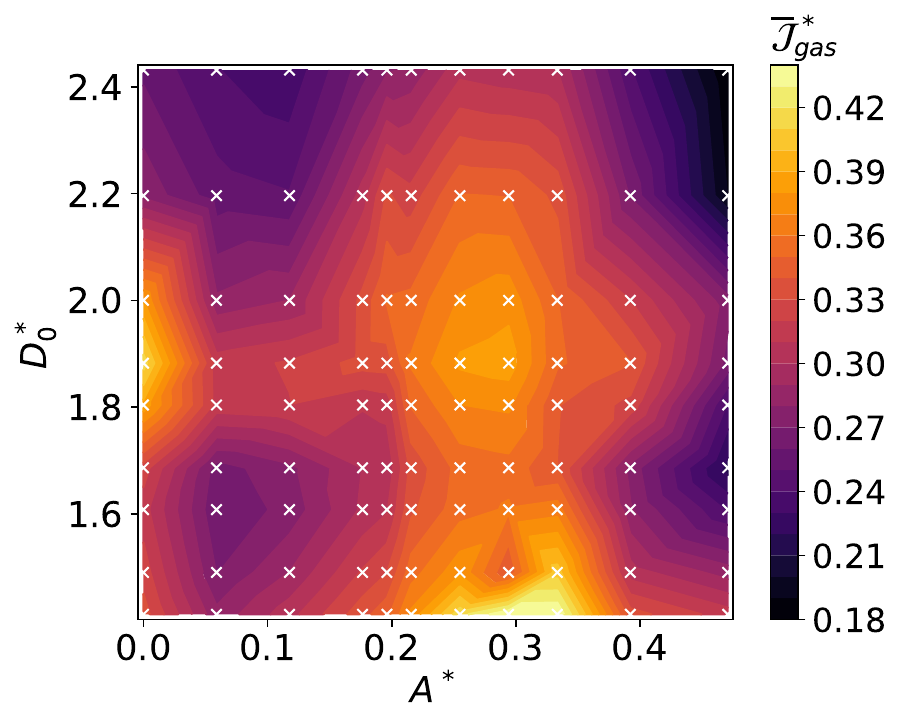}
    \caption{$a^*=4.8$, $\omega^*=4.1 \times 10^3$}
  \end{subfigure}

  \begin{subfigure}{.3\linewidth}
    \centering
    \includegraphics[width=\linewidth]{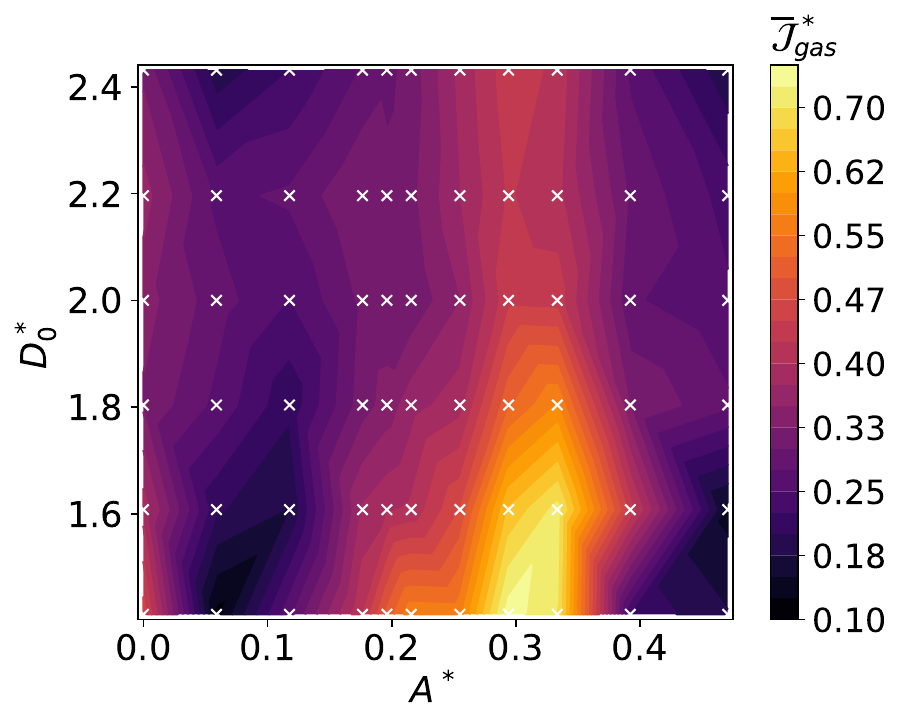}
    \caption{$a^*=1.0$, $\omega^*=2.0 \times 10^3$}
  \end{subfigure}%
  \begin{subfigure}{.3\linewidth}
    \centering
    \includegraphics[width=\linewidth]{A-D0_gasout_mean_rr5e-4_acc5e-5_2D_nd.pdf}
    \caption{$a^*=2.4$, $\omega^*=2.0 \times 10^3$}
  \end{subfigure}%
  \begin{subfigure}{.3\linewidth}
    \centering
    \includegraphics[width=\linewidth]{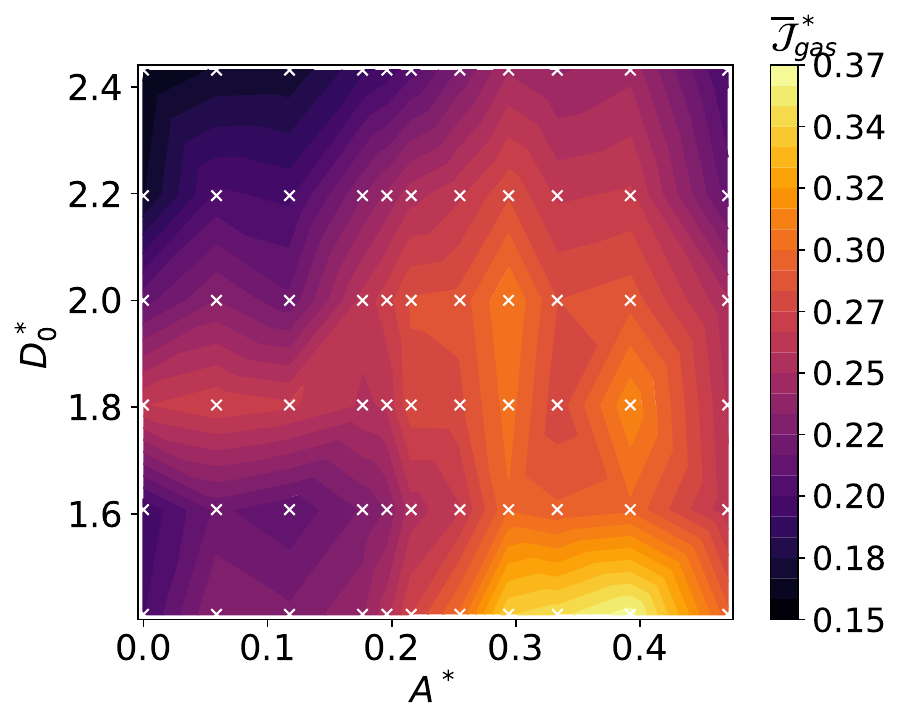}
    \caption{$a^*=4.8$, $\omega^*=2.0 \times 10^3$}
  \end{subfigure}

  \begin{subfigure}{.3\linewidth}
    \centering
    \includegraphics[width=\linewidth]{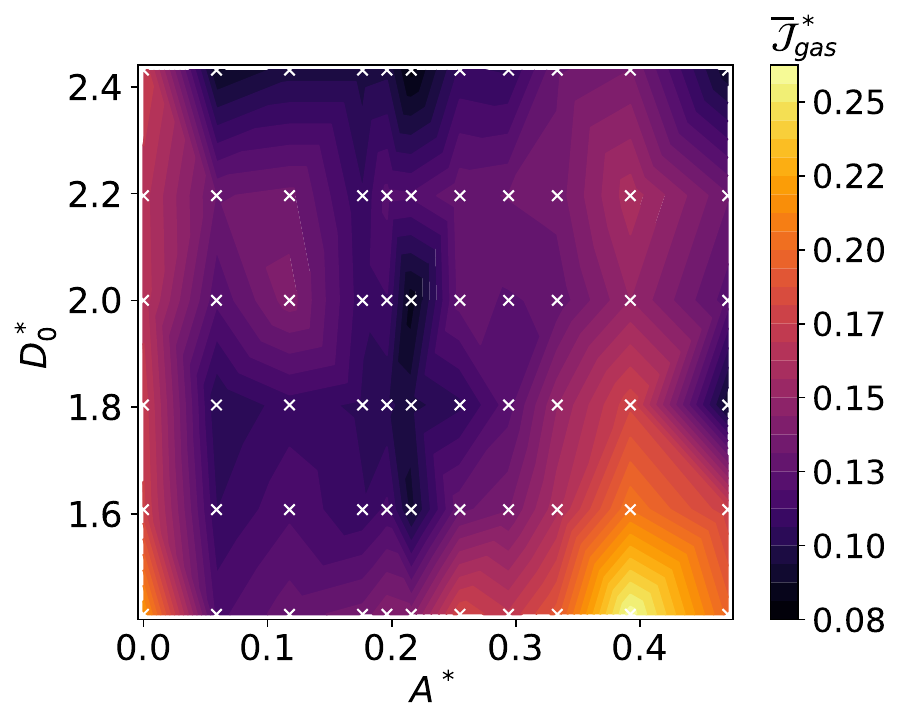}
    \caption{$a^*=1.0$, $\omega^*=4.1 \times 10^2$}
  \end{subfigure}%
  \begin{subfigure}{.3\linewidth}
    \centering
    \includegraphics[width=\linewidth]{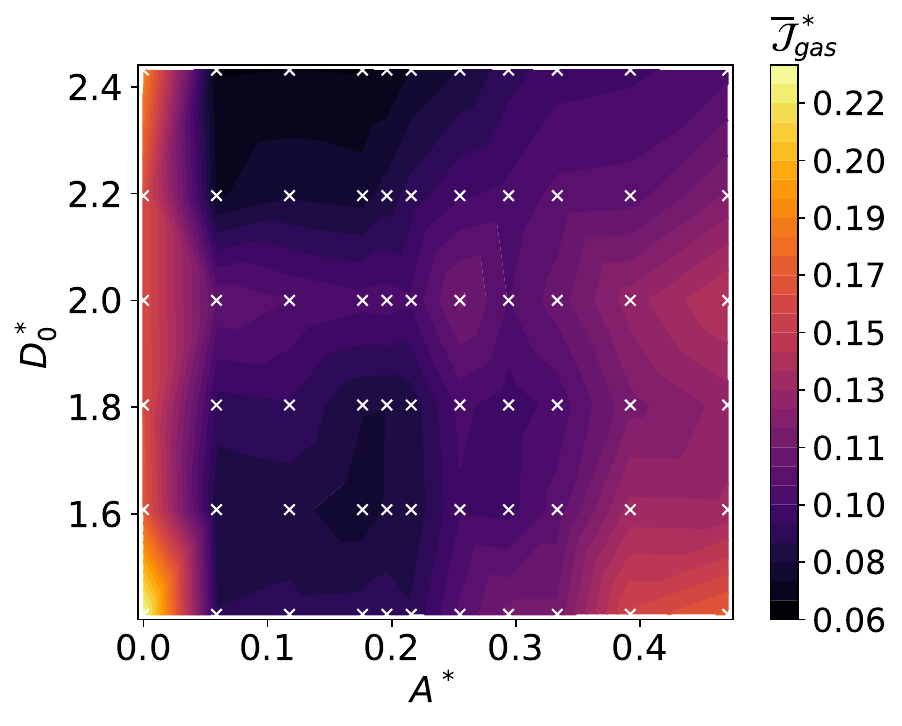}
    \caption{$a^*=2.4$, $\omega^*=4.1 \times 10^2$}
  \end{subfigure}%
  \begin{subfigure}{.3\linewidth}
    \centering
    \includegraphics[width=\linewidth]{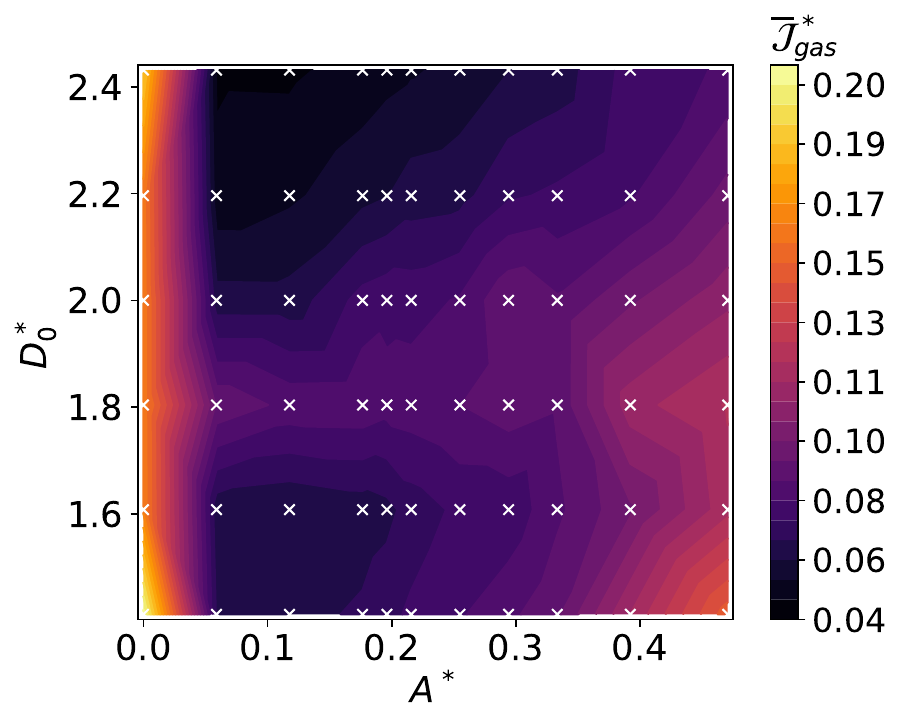}
    \caption{$a^*=4.8$, $\omega^*=4.1 \times 10^2$}
  \end{subfigure}

  \caption{\patchcaption Visualization of the parameter space, where the color coding indicates the corresponding gas output $\overline{J}_gas^*$.}
  \label{fig:phase-space-gasout}
\end{figure*}

To conclude, we broaden the perspective by providing a representative overview of the four-dimensional parameter space spanned by amplitude $A^*$, diameter $D_0^*$, applied acceleration $a^*$ and reaction rate $\omega^*$ as depicted in Fig.~\ref{fig:phase-space-gasout}. 
 
It confirms the trends which we discussed in the previous sections across a wide spectrum of parameters. The overall gas output can be enhanced by reducing the applied acceleration or by increasing the reaction rate, provided channel clogging is avoided. Specifically, we observe channel clogging for $Ca \lesssim 0.1$. In most cases a small channel diameter is superior to large channel diameters and for most of the parameter space the efficiency can be improved by choosing an optimal corrugation amplitude around $A^*\approx 1/3$. Conversely, a flat channel geometry ($A^*=0$) proves beneficial for high accelerations in combination with low reaction rates (Fig.~\ref{fig:phase-space-gasout} h-i). The flat channel geometry stands also out as particularly suited for low accelerations coupled with large reaction rates, as alternative configurations tend to result in channel clogging. Notably, this latter parameter combination yields the highest gas output across all simulations (see Tab.~\ref{tab:top10}).

\begin{table}[h]
\centering
\begin{tabular}{|c|c|c|c|c|c|c|}
\hline
 & $\overline{J}_{gas}^*$ & $A^*$ & $\text{D}_0^*$ & $\omega^*$ & $a^*$  \\
\hline
1 & 1.000 & 0.0 & 1.6 & $4.0 \times 10^3$ & 0.5 \\
2 & 0.928 & 0.0 & 1.2 & $4.0 \times 10^3$ & 1.0 \\
3 & 0.906 & 0.0 & 1.4 & $4.0 \times 10^3$ & 0.5 \\
4 & 0.897 & 0.0 & 1.4 & $3.6 \times 10^3$ & 0.5 \\
5 & 0.878 & 0.0 & 1.6 & $3.6 \times 10^3$ & 0.5 \\
6 & 0.871 & 0.3 & 1.4 & $4.0 \times 10^3$ & 1.0 \\
7 & 0.869 & 0.3 & 1.6 & $4.0 \times 10^3$ & 1.0 \\
8 & 0.859 & 0.3 & 1.4 & $4.0 \times 10^3$ & 1.0 \\
9 & 0.848 & 0.3 & 1.2 & $4.0 \times 10^3$ & 2.4 \\
10 & 0.841 & 0.3 & 1.4 & $4.0 \times 10^3$ & 1.4 \\
\hline
\end{tabular}
\caption{\patchcaption Most efficient parameter combinations based on a sampling of the parameter space with more than $2700$ simulations.}
\label{tab:top10}
\end{table}

\section{Conclusions and Outlook}

In our simulations, we investigated how the yield of gas output can be increased by optimizing the geometry of the porous electrode structure under different flow velocities and reaction rates. To identify beneficial parameter combinations, we sampled the relevant parameter space with more than $2700$ simulations. Our results affirm the presence of advantageous parameter combinations that significantly enhance gas output.

In many cases, gas output can be augmented by increasing the channel corrugation amplitude $A$, thereby leveraging a higher number of reacting sites. However, this trend is effective only up to a point where the corrugation becomes so pronounced that gas starts to accumulate in the pores, rendering reactive sites inactive. Simultaneously, the amplitudes then extend so far into the channel that permeability deteriorates and the efficient advection of gas bubbles is impeded. For very low accelerations or low reaction rates, a flat channel geometry is found to be optimal. 

Further, it generally proves to be beneficial to reduce the flow velocity within the channel, while avoiding complete clogging. This promotes the formation of larger bubbles and capitalizes on the coalescence-induced drop detachment effect. Consequently, the number of active reactive sites is increased and bubbles are directed toward the center of the channel, where higher flow velocities enhance efficient gas bubble transport.

As long as clogging is prevented, higher reaction rates typically result in increased gas output. This is attributed not only to the increased gas production, but also to the fact that higher reaction rates lead to the formation of larger gas bubbles. In analogy to the reasoning for lower accelerations, these larger bubbles are more efficient due to coalescence-induced bubble detachment. 

Despite employing a simplified 2D model, we still encounter the challenge of dealing with an extensive number of possible parameter combinations, resulting in substantial computational cost in exploring the relevant parameter space. Hence, a preemptive exploration of the high-dimensional parameter space was impractical. Future efforts could focus on further exploring optimal parameter combinations, employing techniques such as gradient descent methods or neural networks. Another parameter of great interest for future studies is the wettability of the porous structure, which significantly influences the detachment of gas bubbles. Furthermore, investigating the influence of additional geometrical parameters would allow, e.g., to understand the influence of the tortuosity on the transport in realistic morphologies. 

\begin{acknowledgements}
We acknowledge funding by the Deutsche Forschungsgemeinschaft (DFG, German Research Foundation)—Project-ID 431791331—SFB 1452. Furthermore, we acknowledge the Helmholtz Association of German Research Centers (HGF) and the Federal Ministry of Education and Research (BMBF), Germany for supporting the Innovation Pool project “Solar $H_2$: Highly Pure and Compressed”. We thank the J\"ulich Supercomputing Center for providing the required supercomputing resources and acknowledge Marcello Sega, Tilman Richter, Johannes Hielscher, and Manuel Hegelheimer for fruitful discussions and technical support.

The data that support the findings of this study are openly available at
\href{http://doi.org/10.5281/zenodo.10723216}{doi:10.5281/zenodo.10723216}. 
\end{acknowledgements}

\bibliography{bib.bib}

\end{document}